\documentclass[aps,prd,12point,twocolumn,nofootinbib,showpacs,superscriptaddress]{revtex4-2}
\def\theequation{\arabic{section}.\arabic{equation}}
\usepackage{psfrag}
\usepackage{mathrsfs}
\usepackage{graphicx}
\usepackage{amssymb, bm}
\usepackage{amsmath, amsthm}
\usepackage{epstopdf}
\usepackage{hyperref}
\usepackage{enumerate}
\usepackage{longtable}

\usepackage{amsfonts}
\usepackage[T1]{fontenc}
\usepackage{bm}
\usepackage{amsmath}  
\usepackage{graphicx} 
\usepackage{color}
\usepackage{appendix}

\newcommand{\blue}{\color{blue}}

\newcommand{\be}{\begin{equation}}
\newcommand{\ee}{\end{equation}}
\renewcommand{\d}{\mbox{${\rm d}$}}

\begin{document}
\def\theequation{\arabic{section}.\arabic{equation}} 

\title{The curious case of the Buchdahl-Land-Sultana-Wyman-Iba\~nez-Sanz 
spacetime}

\author{Valerio Faraoni}
\email[]{vfaraoni@ubishops.ca}
\affiliation{Department of Physics \& Astronomy, Bishop's University, 
2600 College Street, Sherbrooke, Qu\'ebec, 
Canada J1M~1Z7}

\author{Sonia Jose}
\email[]{soniatjose96@gmail.com}
\affiliation{Department of Physics \& Astronomy, Bishop's University, 
2600 College Street, Sherbrooke, Qu\'ebec, 
Canada J1M~1Z7}

\author{Alexandre Leblanc}
\email[]{Alexandre.Leblanc3@usherbrooke.ca}
\affiliation{Department of Physics, 
Universit\'e de Sherbrooke, 2500 Boulevard de 
l’Universit\'e, Sherbrooke, Qu\'ebec, 
Canada}



\begin{abstract}

We revisit Wyman's ``other'' scalar field solution of the Einstein 
equations and its Sultana generalization to positive cosmological 
constant, which has a finite 3-space and corresponds to a special case of 
a stiff fluid solution proposed by Buchdahl and Land and, later, by 
Iba\~nez and Sanz to model relativistic stars. However, there is a hidden 
cosmological constant and the peculiar geometry prevents the use of this 
spacetime to model relativistic stars.

\end{abstract}


\maketitle

\section{Introduction}
\label{sec:1}
\setcounter{equation}{0}

An analytical solution of the Einstein field equations of general 
relativity (GR) that is static and spherically symmetric appears in 
two different contexts that are apparently unrelated. In the first 
context, it is a non-asymptotically flat solution of the Einstein 
equations with a free scalar field as a source and with zero cosmological 
constant $\Lambda$, and it was discovered by Wyman in 1981 \cite{Wyman81}. 
This is sometimes called Wyman's ``other'' solution to distinguish it from 
the more well known solution found by Fisher \cite{Fisher:1948yn} and 
rediscovered many times, which in the literature goes by the names 
Fisher-Bergmann-Leipnik-Janis-Newman-Winicour-Buchdahl-Wyman \cite{BL57, 
JNW68, Buchdahl72,Wyman81} (see also Ref.~\cite{Virbhadra:1997ie}) 
 and is the general solution of the $\Lambda=0$ Einstein 
equations which is static, spherically symmetric, asymptotically flat, and 
is sourced by a free scalar field \cite{Fisher:1948yn, BL57, JNW68, 
Buchdahl72}, see \cite{Faraoni:2021nhi} for a recent review.

In the second context, Wyman's ``other'' solution is a special case of 
geometries proposed to describe the interior of a relativistic star by 
Iba\~nez \& Sanz \cite{IbanezSanz} and corresponding to the stiff equation 
of state. Sultana generalized Wyman's ``other'' solution by including a 
positive cosmological constant \cite{Sultana:2015lja}, obtaining a special 
case of another class of perfect fluid solutions found by Iba\~nez \& 
Sanz. More 
precisely, Wyman's ``other'' metric is a special case of a perfect fluid 
geometry found in 1982 by Iba\~nez \& Sanz \cite{IbanezSanz} and in 1968 
by Buchdahl \& Land \cite{BuchdahlLand}, which is itself a special case of 
the Tolman~IV class of GR solutions introduced in 1939 \cite{Tolman39, 
Delgaty:1998uy, Stephanietal}. We summarize below the rather convoluted 
history of the GR solution that is the subject of this work and that we 
call Buchdahl-Land-Sultana-Wyman-Iba\~nez-Sanz (in short, BLSWIS) 
solution.

The BLSWIS metric is contained as a special limit in the Buchdahl \& 
Land's \cite{BuchdahlLand} 1968 stiff fluid solution of the Einstein 
equations with vanishing cosmological constant but pressure 
\be
P= \rho-\rho_0 \,
\ee
where $\rho$ is the fluid energy density and $\rho_0$ is a constant. This 
equation of state was meant \cite{BuchdahlLand} to generalize the 
Schwarzschild interior solution for an incompressible fluid 
\cite{Waldbook} but, apparently unbeknownst to these authors, in practice 
it reintroduces $\Lambda$ into the scenario. The general Buchdahl-Land 
solution is itself a special case of the 1939 Tolman~IV class of solutions  
\cite{Tolman39} describing the interior of a perfect fluid ball with 
$\Lambda$ \cite{Delgaty:1998uy}. As most authors solving for relativistic 
stellar interiors, Buchdahl \& Land \cite{BuchdahlLand} did not match the 
fluid solution to an exterior, the implicit assumption in this literature 
being that the interior is matched with a Schwarzschild exterior at the 
star 
boundary, where the pressure vanishes \cite{Stephanietal, Delgaty:1998uy}.

Wyman's ``other'' solution was found 
in 1981 \cite{Wyman81} as a free scalar field solution of the $\Lambda=0$ 
Einstein equations extending to infinite radius and non-asymptotically 
flat\footnote{The Wyman geometry, but with a different scalar field, is a 
special case of the spacetime reported as a solution of a scalar-tensor 
gravity with power-law potential in Ref.~\cite{Carloni:2013iip} without 
making the connection with \cite{BuchdahlLand,Wyman81,IbanezSanz}. 
However, the geometry and scalar field proposed in~\cite{Carloni:2013iip} 
fail to satisfy the corresponding field equations.} (this work 
\cite{Wyman81} by Wyman is better known because it rediscovered the 
different Fisher-Janis-Newman-Winicour-Buchdahl-Wyman solution and gave it 
in its most general form \cite{Faraoni:2021nhi}). This is a very different 
context from stellar models. In 2015, Sultana \cite{Sultana:2015lja} 
generalized Wyman's ``other'' 
scalar field solution \cite{Wyman81} to the case in which a cosmological 
constant $\Lambda>0$ appears in the Einstein equations. Sultana was 
well aware 
of the fact that the geometry thus obtained is a special case of the 
Iba\~nez \& Sanz 
 solution.\footnote{Sultana's generalization was later used to 
generate an exact solution of Brans-Dicke theory \cite{Brans:1961sx} and 
of $f( {\cal R})= {\cal R}^2$ gravity \cite{Banijamali:2019gry}.} We will 
refer to the scalar field solution of \cite{Sultana:2015lja} as the 
Sultana-Wyman solution (this is the same as the BLSWIS one, but the name 
``Sultana-Wyman'' is a reminder of the fact that the spacetime is sourced 
by $\Lambda$ and by a homogeneous scalar field).

The BLSWIS geometry is contained, as a special case, in the more general 
perfect fluid solution of the Einstein equations with equation of state 
$P=w\rho$, $w=$~const., and $0 < w \leq 1$ found by Iba\~nez \& Sanz in 
1982 \cite{IbanezSanz}. These authors remark that this special case 
had been previously found by Buchdahl \& Land 
\cite{BuchdahlLand}\footnote{Iba\~nez \& Sanz \cite{IbanezSanz} also do 
not match this interior solution to an exterior one.} but they were  
unaware of Wyman's (then recent) paper and they did not realize that, in 
their special case $w=1$, they were introducing the cosmological constant 
even though their field equations are initially declared to have 
$\Lambda=0$ \cite{IbanezSanz}.

As is common in the history of analytical solutions of the Einstein 
equations \cite{Stephanietal}, the same spacetime has been discovered and 
reinterpreted more than once and it is time to introduce some order in the 
relevant literature spanning many decades. This is the purpose of the 
present work, where we revisit the BLSWIS spacetime and compare, as much 
as possible, the two different points of view, 
{\em i.e.}, perfect fluid without scalar field versus scalar field 
solution with $\Lambda>0$. In particular, the boundary conditions for the 
Einstein equations need to be discussed and make stellar models  
based on the BLSWIS geometry unappealing from the physical point of view,  
or even impossible.

We follow the notation of Ref.~\cite{Waldbook}: the metric signature is 
${}{-}{+}{+}{+}$ and we use units in which Newton's constant $G$ and the 
speed of light $c$ are unity, but we occasionally restore $G$ to compare 
with previous literature. $\Lambda$ denotes the cosmological constant and 
$\kappa \equiv 8\pi G$.

\section{The Wyman and Sultana-Wyman scalar field solutions of the 
Einstein equations}
\label{sec:2}
\setcounter{equation}{0}

The Einstein equations sourced by a minimally coupled, free and massless 
scalar field $ \phi $ are
\begin{eqnarray}
\mathcal{R}_{ab}-\frac{1}{2} \, g_{ab} \mathcal{R} +\Lambda g_{ab}
&=&\kappa\Big( \nabla_{a} \phi \nabla_{b} \phi 
-\frac{1}{2} \, g_{ab} \nabla^{c} \phi
\nabla_c \phi \Big) \,, \label{Einstein}\nonumber\\
&&
\end{eqnarray}
\begin{eqnarray}
\Box \phi =0 \,, \label{KleinGordon}
\end{eqnarray}
where ${\cal R}_{ab}$, ${\cal R}$, and $g_{ab}$ are the Ricci tensor, 
Ricci 
scalar, and metric tensor, respectively, while $\nabla_a $ is the 
covariant derivative associated with $g_{ab}$ and  $ \Box \equiv g^{ab} 
\nabla_a \nabla_b$ is the curved 
space d'Alembertian.

The general static, spherically symmetric, and asymptotically flat 
solution of these equations for $\Lambda=0$ is the well known 
Fisher solution \cite{Fisher:1948yn, BL57, JNW68, Buchdahl72, Wyman81, 
Virbhadra:1997ie} (see the 
recent review \cite{Faraoni:2021nhi} for a discussion of this and other 
spherical solutions). Under the assumption that the 
matter field $\phi$ depends only on the radial coordinate, the unique 
static, spherical, and asymptotically flat solution was found by 
Fisher \cite{Fisher:1948yn} and later rediscovered, in other coordinates 
or in 
other forms, by Bergmann \& Leipnik \cite{BL57}, Janis, Newman \& 
Winicour \cite{JNW68}, Buchdahl \cite{Buchdahl72}, and finally by Wyman 
\cite{Wyman81}, who wrote the most general form of this 
solution. Wyman proposed another family of solutions for 
$\Lambda=0$  (generalized by Varela \cite{Varela:1987td} to the case 
$\Lambda \neq 
0$) corresponding to 
spherically symmetric and static geometry and with scalar field depending 
only on time, $\phi=\phi(t)$. In general, this class of solutions is 
expressed by power series and is not useful for practical calculations, 
but one of them (again, for $\Lambda=0$) is particularly simple 
\cite{Wyman81}:
\begin{eqnarray}
d s^2 =-\kappa r^2 dt^2+2dr^2 +r^2 d\Omega^2_{(2)} \,, \label{act} 
\end{eqnarray}
\begin{eqnarray}\label{phi}
\phi(t)=\phi_0 \, t \,,
\end{eqnarray}
where $ d\Omega^{2}_{(2)}=d\vartheta^{2}+\sin^{2}\vartheta \, 
d\varphi^{2}$ is 
the line element on the unit 2-sphere and $\phi_0$ is a dimensionless 
constant. We refer to this solution as Wyman's ``other'' solution.  It is 
a special case 
(for $w=1$) of the ``scaling solution'' published a year later by Iba\~nez 
\& Sanz 
\cite{IbanezSanz} for a perfect fluid with equation of 
state\footnote{Iba\~nez \& Sanz use units in which $\kappa=1$. Therefore,  
their Eqs.~(15)  for the energy density and pressure differ from our 
Eqs.~(\ref{rhow}), (\ref{Pw}) by  a factor $8\pi$ in the denominator.}   
$P=w\rho$
\be
ds^2 =-r^{ \frac{4w}{1+w} } dt^2 + \frac{ w^2+6w+1}{(w+1)^2} \, dr^2 +r^2 
d\Omega_{(2)}^2 \,. \label{scalingsolution}
\ee
In spite of the fact that the energy density 
\cite{IbanezSanz}
\be
\rho(r)  = \frac{w}{2\pi (w^2+6w+1) r^2} 
\ee
and the pressure $ P(r) = w\rho(r) $ are singular at $r=0$, this solution 
is usually regarded as possessing regions that are realistic 
approximations to the bulk of a star on the verge of collapsing 
\cite{HTWW,Visser:2002ww,Jowsey:2021ixg,Stephanietal}.

Sultana \cite{Sultana:2015lja} has generalized Wyman's ``other'' solution to 
include a positive cosmological constant $\Lambda$. The scalar 
field remains linear in time as in Eq.~(\ref{phi}) (see  
Appendix~\ref{Appendix:A} for  a discussion),  while the line 
element becomes 
\begin{eqnarray}\label{geo}
ds^2 = -\kappa r^2 dt^2 +\frac{2dr^2 }{1-\frac{2\Lambda
r^2}{3} } +r^2 d\Omega^2_{(2)} 
\end{eqnarray}
(we will refer to this, in conjunction with Eq.~(\ref{phi}) for the 
scalar, as the ``Sultana-Wyman solution''). 
The limit $\Lambda\rightarrow 0$ reproduces  Wyman's ``other'' 
solution~(\ref{act}) and (\ref{phi}). 
Again, the Sultana-Wyman solution is a special case of a family found by 
Iba\~nez \& Sanz \cite{IbanezSanz} with the Heintzmann method 
\cite{Heintzmann69}, which 
generalizes the ``scaling solution'' (\ref{scalingsolution}):
\be
ds^2 =- r^{ \frac{4w}{1+w} } dt^2 + \frac{a}{ 1-Car^{2+b} } \, dr^2 
+r^2 d\Omega_{(2)}^2 \,, \label{urca}
\ee
where 
\begin{eqnarray}
a & = & \frac{ w^2+6w+1}{(w+1)^2} \,,\label{100:cz1}\\
&&\nonumber\\
b &=& \frac{4w(1-w)}{(w+1)(3w+1)} \,,\label{100:cz2}
\end{eqnarray} 
and where $C$ is an arbitrary constant. The corresponding energy density 
and pressure are
\cite{IbanezSanz} 
\begin{eqnarray}
\rho_w(r) &=& \frac{1}{8\pi} \left[ \frac{4w}{ (1+w)^2 a r^2} + C(3+b) r^b 
\right] \,, \label{rhow}\\
&&\nonumber\\
P_w(r) &=& \frac{1}{8\pi} \left[ \frac{4w^2}{(1+w)^2 a r^2} - \frac{C 
(1+5w)}{(1+w)} \,  r^b  \right] , \label{Pw}
\end{eqnarray}
which are singular at $r=0$ (Iba\~nez \& Sanz consider the range of 
equation of state parameters $0<w\leq 1$ and find no solutions for dust 
$w=0$, which would eliminate the divergence in $\rho_w$ and $P_w$ 
\cite{IbanezSanz}). For 
$w=1$, which corresponds to the stiff equation of state of a free scalar 
field, it is $a=2$, $b=0$ and the energy density and pressure become 
\begin{eqnarray}
\rho_1 (r) &=& \frac{1}{8\pi} \left[ \frac{1}{2 r^2} + 3C \right] 
\,,\label{rho1}\\
&&\nonumber\\
P_1(r)  &=& \frac{1}{8\pi} \left[ \frac{1}{2r^2} - 3 C 
\right]  \,.\label{P1}
\end{eqnarray}
It is interesting that Iba\~nez \& Sanz  \cite{IbanezSanz}  do not relate 
their constant $C$ to the cosmological constant in the $w=1, b=0$ case, 
although it is clear that the last term in the right hand 
side of Eq.~(\ref{rho1}) and of Eq.~(\ref{P1}) can be regarded as the 
contribution of a cosmological constant $\Lambda=3C$ to the total energy 
density and pressure, added to those of the free scalar field. Moreover, 
for $w=1$ the line element~(\ref{urca}) generalizes the Wyman solution 
also to the case $\Lambda<0$ (this solution is implicit in Sultana's paper 
\cite{Sultana:2015lja}).

The physical nature of Wyman's ``other'' solution was studied in previous 
papers \cite{BuchdahlLand,IbanezSanz,Banijamali:2019gry, Jowsey:2021ixg} 
and its Sultana generalization was used in~\cite{Banijamali:2019gry} to 
generate a new solution of Brans-Dicke theory with a massive scalar by 
means of a conformal transformation to the Jordan frame (the same geometry 
is a solution of $f({\cal R})={\cal R}^2$ gravity 
\cite{Banijamali:2019gry}). Using the same method, 
Ref.~\cite{Sultana:2015lja} generated new solutions of conformally coupled 
scalar field theory with a Higgs potential.

Let us analyze the physical properties of the Sultana-Wyman 
solution~(\ref{geo}) and~(\ref{phi}) for $\Lambda>0$. The time and radial 
coordinates vary in the range \begin{eqnarray} 
-\infty<t<+\infty,\,\,\,\,\,\,\,\, \;\; 0\leq r<\sqrt{\frac{3}{2\Lambda}}. 
\end{eqnarray}

\subsection{Geometry and radial geodesics}

The Sultana-Wyman geometry described by the line element~(\ref{geo}) is 
static and spherically symmetric. By taking the limit $\Lambda\rightarrow 
0$, one recovers Wyman's ``other'' solution (\ref{act}), (\ref{phi})  
extending to $ 0\leq r < +\infty$. 

In general, if (apparent) horizons are present in a spherically symmetric 
geometry, they are located by the roots of the equation
\be 
g^{ab}\,\nabla_{a}r \nabla_{b}r = 
g^{rr}=0 \,,
\ee  
where $r$ is the areal radius  ({\em e.g.}, \cite{myAHbook}), which is 
always defined in the presence of spherical symmetry. (Furthermore, a single 
root denotes  a black hole or white hole apparent horizon, while a double 
root denotes a wormhole horizon throat \cite{myAHbook}.)  
In the Sultana-Wyman case~(\ref{geo}), this equation has the unique single   
root
\begin{eqnarray}
r_*=\sqrt{\frac{3}{2\Lambda}} 
\end{eqnarray}
which, however, does not correspond to a horizon. To understand this 
situation note that, in spite of the fact that the time direction $t^a 
=\left( \partial/\partial t \right)^a$  is a timelike Killing vector of  
the geometry~(\ref{geo}), its norm 
\be
t_a t^a = -kr^2
\ee
does not change sign anywhere and, unlike what happens for the 
Schwarzschild metric 
or the de Sitter metric, there is no Killing 
horizon here. The 3-dimensional space $t=$~const. is finite and is covered 
by 
the range $0<r\leq \sqrt{\frac{3}{2\Lambda}} $ of the radial coordinate. 
The scalar field and the cosmological constant satisfy the weak and null 
energy conditions and generic strong rigidity arguments lead one to 
exclude event horizons \cite{HawkingEllis, Carter:1969zz, 
Chrusciel:1996bj, Friedrich:1998wq, Racz:1999ne} given the absence of 
Killing horizons.

\begin{figure}
\includegraphics[scale=0.35]{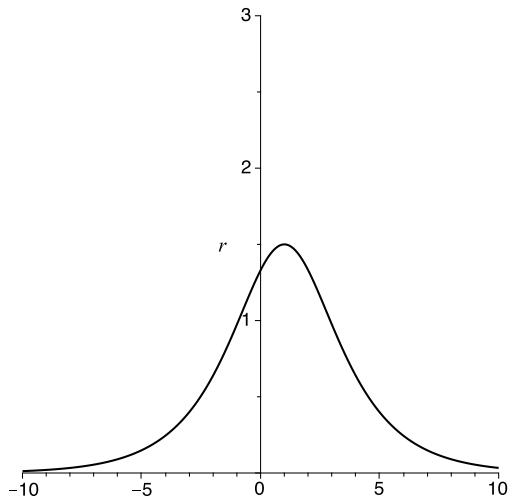}
\caption{\label{fig:3} The radial coordinate $r(\lambda)$ of a 
radial photon (vertical axis) versus the affine parameter $\lambda$ 
(equivalently, the time $t$, on the horizontal axis) for the parameter 
values $\kappa=1, \lambda_0=1$. The region  $\lambda<\lambda_0$ to the 
left of the peak described outgoing photons with $dr/d\lambda>0$, 
while $\lambda>\lambda_0$ describes radial ingoing photons. A photon 
starting near $\lambda=-\infty$ and $r\simeq 0$ takes a very long  
time to reach the turning point (represented by the peak 
$r_\mathrm{max}=\sqrt{\frac{3}{2\Lambda}}$ of $r(\lambda))$. From there, 
the radial photon returns toward the origin $r=0$  circling the finite 
3-space, while approaching $r=0$ in an infinite $\lambda$-time.   
} 
\end{figure}

To confirm this property, consider the congruences of outgoing ($+$) and 
ingoing ($-$) radial null 
geodesics with tangents $l^c_{(\pm)}$ and components $l^{\mu}_{(\pm)}= 
\left( l^0, 
l^1, 0,0 \right)$. The normalization $ l_a^{(\pm)} l^a_{(\pm)} =0$ yields
\be
 l^1_{(\pm)} = \pm \sqrt{ \frac{\kappa}{2}} \, r \sqrt{ 
1-\frac{2\Lambda r^2}{3}} \, l^0_{(\pm)}  
\ee
and, since a null vector can be rescaled by a function, we can choose 
$l^0=1$ obtaining
\be
l^{\mu}_{(\pm)} = \Big( 1, \pm \sqrt{ \frac{\kappa}{2} \,  
\left( 1-\frac{2\Lambda r^2}{3}\right)  } \, r ,0, 0 \Big) \,.
\ee
The equation of radial null geodesics can be integrated remembering that 
$ l^{\mu} \equiv dx^{\mu}(\lambda)/d\lambda$, where $\lambda$ is an affine 
parameter along the null geodesics. Then we have 
\begin{eqnarray}
\frac{dt}{d\lambda} &=& 1 \,,\\
&&\nonumber\\
\frac{1}{r\sqrt{ 1-2\Lambda r^2/3} } \, \frac{dr}{d\lambda} &=& \pm \sqrt{ 
\frac{ \kappa}{2} } \,,
\end{eqnarray} 
and then
\begin{eqnarray}
&& t(\lambda) = \lambda-\lambda_0 \,,\\
&&\nonumber\\
&& - \mbox{arctanh} \left( \sqrt{1-\frac{2\Lambda r^2}{3}} \right) = \pm 
\sqrt{\frac{\kappa}{2}} \left( \lambda -\lambda_0 \right) \,.\label{mmm}
\end{eqnarray}
Simple manipulations of Eq.~(\ref{mmm}) yield
\be
r(\lambda) = \sqrt{ \frac{3}{2\Lambda} } \frac{1}{\cosh \left[ 
\sqrt{ \frac{\kappa}{2} } \left( \lambda-\lambda_0 \right) \right] } \,,
\ee
see Fig.~\ref{fig:3}, which shows that radial null geodesics can never 
reach 
radii larger than 
$ \sqrt{ \frac{3}{2\Lambda}} $ since $\cosh x \geq 1$. A photon 
at $r=0$ is  an infinite value of the affine 
parameter  $\lambda$ away from the turning point  
$r=\sqrt{\frac{3}{2\Lambda}} $. It takes an arbitrarily long time  
 $t \sim 
\lambda$ for a photon arbitrarily close to $r=$ to arrive to the turning 
point $r=\sqrt{\frac{3}{2\Lambda}} $. Similarly, a photon 
travelling radially  and starting at  
$r=\sqrt{\frac{3}{2\Lambda}} $ at $\lambda=\lambda_0$ (or at any 
finite radius) takes an infinite $\lambda$-time to reach the 
origin (Fig.~\ref{fig:3}). 

\begin{figure}
\includegraphics[scale=0.35]{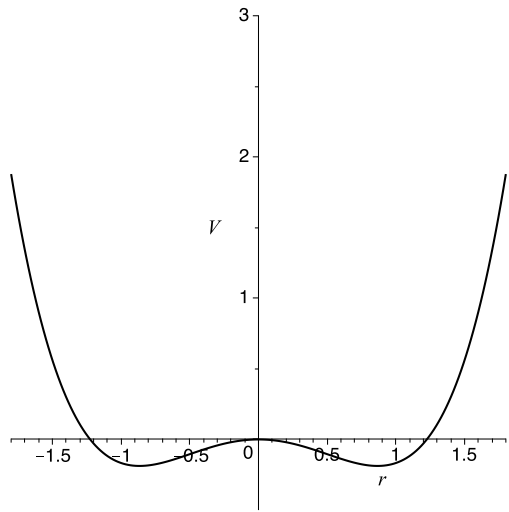}
\caption{\label{fig:1} The potential $V(r) $ (only the region $r\geq 0$ 
is physical and $\kappa$ and $\Lambda$ are set to unity for 
illustration). The motion is confined between $r=0$ and the turning point  
$\sqrt{\frac{3}{2\Lambda}}$ because the total energy is zero.  A radial 
outgoing photon ($dr/d\lambda>0$) starting out arbitrarily close to $r=0$ 
in the far past takes an arbitrarily long time (until 
$\lambda_0$) to reach the turning point $r=\sqrt{ \frac{3}{2\Lambda}}$ and 
then heads again for $r=0$, approaching in an infinite time and circling 
the finite 3-space. $dr/d\lambda $ vanishes as the photon approaches $r=0$ 
or $r=\sqrt{ \frac{3}{2\Lambda} }$. The photon cannot sit in the minimum 
of the potential $V_\mathrm{min}<0$ because the total 
energy is forced to be zero.} 
\end{figure}

This 
dynamics 
can be understood 
by rewriting  
the $l^1$ component of the four-tangent to radial null geodesics as
\be
\frac{dr}{d\lambda}= \pm \sqrt{ \frac{\kappa}{2}} \, 
r\sqrt{1-\frac{2\Lambda r^2}{3}} \,,
\ee
squaring, and dividing by 2, which yields
\be
\frac{1}{2} \left( \frac{dr}{d\lambda} \right)^2 +V(r) =0 \,, \quad \quad
V(r)=\frac{\kappa r^2}{2} \left( \frac{2\Lambda r^2}{3}-1 \right) \,, 
\ee
a formal energy conservation equation for a fictitious particle of unit 
mass and zero total energy in the effective potential $V(r)$. The latter 
intersects the $r$-axis at $r=0, \sqrt{ \frac{3}{2\Lambda}}$ and has a 
negative minimum $V_\mathrm{min}=- \frac{3\kappa}{32\Lambda}$ at $r=\sqrt{ 
\frac{3}{4\Lambda}} $ (Fig.~\ref{fig:1}). ($V(r)$ is an even function, but 
we are only interested in the region $r\geq 0$.)

Since the energy of the fictitious particle representing the radial photon 
is always zero, the motion is confined between $r=0$ and the turning 
point $\sqrt{ \frac{3}{2\Lambda}}$. If the radial photon  
starts near $r=0$, it must do so with nearly zero 
kinetic energy and it takes an infinite amount of $\lambda$-time to reach 
the turning point $r=\sqrt{ 
\frac{3}{2\Lambda}}$ at the end of 3-space. The point $r=0$ is an unstable 
equilibrium point and 
a particle located there has zero energy and remains there. Once the 
radial photon is at the boundary $r=\sqrt{ \frac{3}{2\Lambda}}$ of the 
finite space, it circles it toward the origin $r=0$, but it takes an 
infinite $\lambda$-time to reach it as this photon slows down approaching 
it. The radial photon completes a single cycle of ``oscillation'' between 
$r=0$ and $r=\sqrt{\frac{3}{2\Lambda}}$ in an infinite time, in a manner 
analogous to an overdamped oscillator.

It might appear that there is a stable circular photon orbit at $ 
r=\sqrt{ \frac{3}{4\Lambda}}$, where the potential is minimum, but photons 
cannot stay there because the total energy must be zero and, since 
$V_\mathrm{min}<0$, the positive kinetic energy $ \left( dr/d\lambda 
\right)^2/2 =-V_\mathrm{min}$ moves it away from this radius.

We can calculate the expansions of the congruences of outgoing and 
ingoing radial null geodesics,
\begin{eqnarray}
\theta_{(\pm)} &=& \nabla_c l^c_{(\pm)} \nonumber\\
&&\nonumber\\
&=& 
\partial_{\mu} l^{\mu}_{(\pm)} 
+\Gamma^{\mu}_{\mu\alpha} l^{\alpha}_{(\pm)} \nonumber\\
&&\nonumber\\
&=&  
\partial_t l^0_{(\pm)} + \partial_r l^1_{(\pm)} 
+ \Gamma^{\mu}_{\mu 0} l^0_{(\pm)}
+ \Gamma^{\mu}_{\mu 1} l^1_{(\pm)} \,,\nonumber\\
&&
\end{eqnarray}
where $\Gamma^{\mu}_{\alpha\beta}$ denote the Christoffel symbols. Using 
\be
\Gamma^{\mu}_{\mu 0}=0 \,, \quad\quad 
\Gamma^{\mu}_{\mu 1} = \frac{3-4\Lambda r^2/3}{ r\left( 1-2\Lambda r^2/3 
\right)} \,,
\ee
one obtains
\be
\theta_{(\pm)} = \pm 2 \sqrt{ 2\kappa} \, \sqrt{1-\frac{2\Lambda r^2}{3} } 
\,.
\ee  
In the limit $r\rightarrow \sqrt{\frac{3}{2\Lambda}} $ both expansions 
vanish. This anomalous behaviour does not characterize a horizon (at 
which one of the expansions vanishes and the other does not), but 
signals the fact that 3-space ends at $r=\sqrt{ \frac{3}{2\Lambda}}$ 
(more on this below). 

Consider now outgoing/ingoing radial timelike geodesics with four-tangents 
\be
p^{\mu}_{(\pm)} = mu^{\mu}_{(\pm)} = m \, \frac{d x^{\mu}}{d\tau}\Big|_{(\pm)} 
 = \left( p^0, p^1_{(\pm)} , 0, 0 \right)\,,
\ee
where $m$ is the mass of a test particle of four-velocity $u^a_{(\pm)}  $ 
and $\tau$ 
is the proper time along the timelike geodesic. The timelike Killing 
vector $t^a$ guarantees conservation of energy along each geodesic:
\be
p_a^{(\pm)}  t^a = -E = \mbox{const.} 
\ee
yielding
\be
u^0 = \frac{\bar{E}}{\kappa r^2} \,,
\ee
where $\bar{E} \equiv E/m$ is the (constant) energy per unit mass. Then the 
normalization $u_a u^a =-1$ gives
\be
u^{\mu}_{(\pm)} = \Big( \frac{ \bar{E} }{\kappa r^2} , \pm \sqrt{ 
\frac{1}{2} 
\left( \frac{ \bar{E}^2 }{\kappa r^2} -1 \right) 
\left( 1 -  \frac{2\Lambda r^2}{3}  \right) }, 0,0 \Big) \,.
\ee
The particle is at rest if either $r=\bar{E}/\sqrt{\kappa}$ (in which case 
$ u^0=1$ and $u^1=0$), or if $r=\sqrt{\frac{3}{2\Lambda}}$ (in which case 
the particle is as far from the origin as possible). 

The coordinate radial velocities of outgoing/ingoing massive test 
particles are
\begin{eqnarray}
\frac{dr}{dt} &=& \frac{dr}{d\tau} \, \frac{d\tau}{dt} = \frac{u^1_{(\pm)} 
}{u^0 }  \nonumber\\
&&\nonumber\\
&=&  \pm  
\sqrt{ \frac{ \left( \bar{E}^2 -\kappa r^2 \right)}{2\bar{E} }   
\left(  1-\frac{2\Lambda r^2}{3} \right) } \,,
\end{eqnarray}
which vanish in the limit $r\rightarrow \sqrt{\frac{3}{2\Lambda}}$ (while 
$u^0 \neq 0$, of course): at this radius, particles do not move either 
outwards or inwards, which would 
not happen at a horizon where only motion in one direction is forbidden 
(outward for a black hole horizon, inward for a cosmological or white 
hole horizon).

\begin{figure}
\includegraphics[scale=0.35]{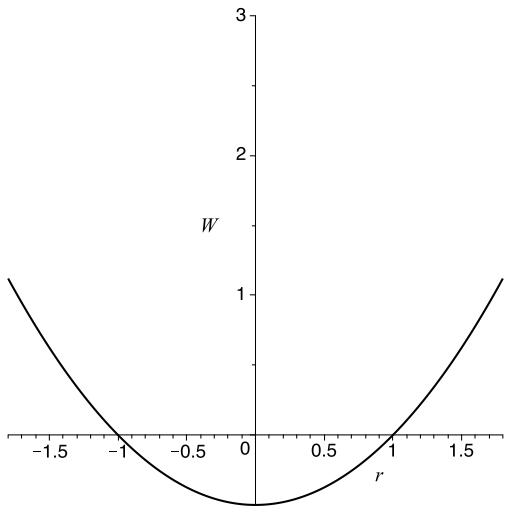}
\caption{\label{fig:2} The harmonic oscillator potential $W(r)$ (only the 
region $r\geq 0$ is physical and $K=1$ for illustration).  A radial photon 
oscillates between $r=0$ and $r=1/\sqrt{K}$, going around the 
finite hyperspherical 3-space again and again, each ``oscillation'' 
taking a finite $\lambda$-time.} 
\end{figure}

It is useful to compare the Sultana-Wyman geometry with the Einstein 
static universe, which has line element
\be
ds^2 = -dt^2 +
\frac{dr^2}{1-Kr^2} +r^2 d\Omega_{(2)}^2 \,,
\ee
with constant curvature index $K>0$ and finite 3-spaces of constant 
time and radial coordinate spanning the finite range $0\leq r\leq 
1/\sqrt{K}$.  Naively, since this 
metric is spherically symmetric and $r$ is the areal radius, a search for 
horizons  with the equation 
\be
\nabla^c r \nabla_c r = g^{rr}= 1-Kr^2=0
\ee
would yield the unique positive single root $r=1/\sqrt{K}$, but we know 
better. 
This is not a horizon, the norm of the timelike Killing vector $\left( 
\partial/\partial t \right)^a $ is always $-1$ and does not change sign 
anywhere,  and we 
expect  the expansions $\theta_{(\pm)}$ 
of radial null geodesics to exhibit pathological behaviour at 
$r=1/\sqrt{K}$. This is indeed the case. Let these geodesics 
have tangents $l^a_{(\pm)}$, then the normalization $ l^a_{(\pm)} 
l_a^{(\pm)} =0 $ yields $ 
l^1_{(\pm)} = \pm \, l^0 \sqrt{1-Kr^2} $ and, choosing again $l^0=1$, one 
has 
\be
l^{\mu}_{(\pm)} = \Big( 1, \pm \sqrt{1-Kr^2} , 0, 0 \Big) \,.
\ee
The geodesic equations 
\be
\frac{dt}{d\lambda} =1 \,, \quad\quad \frac{dr}{d\lambda}=\pm 
\sqrt{1-Kr^2} \,,
\ee
are easily integrated to $ t(\lambda)= \lambda-\lambda_0$ (where 
$\lambda_0$ is an integration constant) and
\be
\frac{\arcsin \left( \sqrt{K}\, r\right) }{\sqrt{K}}= \pm \left( 
\lambda-\lambda_0 \right) \,.
\ee
The last equation gives
\be
r(\lambda) = \pm \frac{1}{\sqrt{K}} \, \sin \left[ \sqrt{K} \left( 
\lambda-\lambda_0 \right) \right] \,,\label{eq:sine}
\ee
where the sign of the right hand side is chosen so that $r(\lambda)$ 
remains non-negative. The periodicity shows that a radial photon keeps 
circling the finite 
3-space along the same spatial curve on the 3-sphere. We can rewrite the  
equation $l^1_{(\pm)}=dr/d\lambda = \pm \sqrt{ 1-Kr^2}$ as
\be
\frac{1}{2} \left( \frac{dr}{d\lambda} \right)^2 +W(r)=0 \,, \quad\quad 
W(r) =\frac{1}{2} \left(Kr^2-1 \right) \,.
\ee
The potential $W(r)$ is that of a simple harmonic oscillator with the 
origin of 
the energy shifted (Fig.~\ref{fig:2}), which intersects the $r$-axis at 
$r=1/\sqrt{K}$, a turning point where the kinetic energy 
vanishes.

Since the total effective energy is zero, the motion is confined between 
$r=0$ and the turning point $r=1/\sqrt{K}$. Radial photons in this finite 
spacetime ``oscillate'' between $r=0$ and $r=\sqrt{ \frac{3}{2\Lambda}}$, 
which physically means that they keep going around the spherical 3-space, 
as described by the periodic solution~(\ref{eq:sine}). There are no stable 
or unstable circular orbits, except for the degenerate one at $r=0$.

The expansions of the radial null geodesic congruences are 
\begin{eqnarray}
\theta_{(\pm)} &=&  \nabla_a l^a_{(\pm)} = \partial_{\mu} l^{\mu}_{(\pm)} 
+ 
\Gamma^{\mu}_{\mu\alpha} l^{\alpha}_{(\pm)} \nonumber\\
&&\nonumber\\
& = &  \pm \partial_r \sqrt{1-Kr^2} 
+\Gamma^{\mu}_{\mu 0} + 
\Gamma^{\mu}_{\mu 1} \sqrt{1-Kr^2} \,.\nonumber\\
&&
\end{eqnarray}
Using 
\be
\Gamma^{\mu}_{\mu 0} = 0 \,, \quad\quad 
\Gamma^{\mu}_{\mu 1} = \frac{2}{r} -Kr \left( 1-Kr^2 \right)  \,, 
\ee
one obtains
\begin{eqnarray}
\theta_{(\pm)} &=&  \pm \frac{1}{ \sqrt{1-Kr^2} } \nonumber\\
&&\nonumber\\
&\, & \times 
 \left\{ -Kr 
+\left(1-Kr^2\right) \left[ \frac{2}{r} -Kr\left(1-Kr^2\right) \right] 
\right\} \,,\nonumber\\
&&
\end{eqnarray}
which correctly reduce to $\pm 2/r$ in the degenerate Minkowski case 
$K=0$. In the 
limit $r\rightarrow 1/\sqrt{K} $, we have $\theta_{(\pm)} \rightarrow \mp 
\infty$, signaling the fact that there is no horizon at this radius, but 
the 3-space is finite instead. 

One can consider also radial timelike geodesics parametrized by the proper 
time 
$\tau$. The timelike Killing vector $t^a = \left( \partial /\partial t 
\right)^a$ with unit norm gives energy conservation along each such 
geodesic: $p_c t^c = -E=$~const. yields
\be
u^0 = \bar{E} \equiv \frac{E}{m} 
\ee
and the normalization $u_c u^c=-1$ then gives
\be
u^{\mu}_{(\pm)} = \Big( \bar{E}, \pm \sqrt{ \left( \bar{E}^2-1\right) \left( 
1-Kr^2\right)}, 0, 0 \Big) \,.
\ee
Radial motion stops either if $\bar{E}=1$ (in which case $u^0=1$)  or if 
$r=1/\sqrt{K}$, where the particle 
is as far away from the origin as possible in the Einstein static universe. 
It is easy to 
integrate the timelike geodesic equation, obtaining
\begin{eqnarray}
t(\tau) &=& \bar{E} \tau +t_0 \,,\\
&&\nonumber\\
r(\tau) &=& \pm \frac{1}{ \sqrt{K}} \sin \left[ \sqrt{ K\left( 
\bar{E}^2-1\right) } \left( \tau-\tau_0 \right) \right] \,,
\end{eqnarray}
(where $t_0$ and $\tau_0$ are integration constants) or, eliminating the 
parameter $\tau$, 
\be
r(t)= \pm \frac{1}{\sqrt{K}} \sin \left[ \sqrt{ K 
\left(1-\frac{1}{\bar{E}^2} \right) } \left(t-t_0\right) \right] \,.
\ee
The radial position of the particle cannot exceed the maximum value 
$1/\sqrt{K}$. 

To conclude, the 3-space of the Sultana-Wyman geometry is finite. Denoting 
with $g^{(3)}$ the determinant of the restriction $g_{ab}^{(3)}$ of the 
spacetime metric $g_{ab}$ to this subspace, its volume is given by
\begin{eqnarray}
V &=& \int d^3 \vec{x} \, \sqrt{g^{(3)} } = \int \frac{ \sqrt{2}\, r^2 
\sin\vartheta}{\sqrt{1-2\Lambda r^2/3} } \, dr\d\vartheta d\varphi 
\nonumber\\
&&\nonumber\\
&=&  4\pi \sqrt{2} \int_0^{ \sqrt{ \frac{3}{2\Lambda} } } dr \, 
\frac{r^2}{\sqrt{1-2\Lambda r^2/3} } \nonumber\\
&&\nonumber\\
&=& \frac{ 4\pi \sqrt{2} }{8} \left[ 
\frac{3\sqrt{6}}{\Lambda^{3/2} } 
\, \arcsin\left( \sqrt{ \frac{2\Lambda}{3} } \, r \right) -2r 
\sqrt{1-\frac{2\Lambda r^2}{3} } \right]_{0}^{ \sqrt{\frac{3}{2\Lambda} } 
}    \nonumber\\
&&\nonumber\\
&=& \frac{ 3\sqrt{3} \, \pi^2}{2\Lambda^{3/2} } \simeq 25.64 
\, \Lambda^{-3/2} \,.
\end{eqnarray}

\subsection{Central singularity}

By computing the Ricci scalar from the Sultana-Wyman line element~(\ref{geo}) 
one obtains $ {\cal R}=4\Lambda-1/r^2 $, while contracting the field 
equations~(\ref{Einstein}) and using Eq.~(\ref{phi}) yields  
\begin{eqnarray}
\mathcal{R} = 4\Lambda + \kappa \, g^{ab}\, \nabla_a \phi\,
\nabla_b \phi 
=4\Lambda-\frac{ \phi_0^2}{r^{2}} \,, 
\end{eqnarray}
which fixes the dimensionless integration constant to $\phi_0=\pm 1 $. 
The Ricci scalar  
diverges as $r\rightarrow 0^{+}$ in both cases $\Lambda=0$ 
(Wyman's ``other'' solution) and 
$\Lambda>0$ (Sultana-Wyman solution). The total ({\em i.e.}, including 
scalar field and cosmological constant) energy density and 
pressure obtained from Eqs.~(\ref{100:cz1})--(\ref{Pw}) for 
$w=1$,
\begin{eqnarray}
\rho(r) &=& \frac{1}{8\pi} \left( \frac{1}{2r^2} +\Lambda \right) \,,\\
&&\nonumber\\
P(r) &=& \frac{1}{8\pi} \left( \frac{1}{2r^2} -\Lambda \right) \,,
\end{eqnarray}
are also singular but the spatially homogeneous 
scalar field is regular everywhere.  The 
Sultana-Wyman solution is interpreted as a scalar field naked central 
singularity embedded in a ``background'' due to the 
cosmological constant.\footnote{The quotation marks are mandatory because, 
due to the non-linearity of the Einstein equations, a metric cannot be 
split into a ``background'' plus a ``deviation'' from it in a covariant 
way, except for (generalized) Kerr-Schild metrics.}

The equation of  a sphere of constant radius $r_0$ is $f(r)= r-r_0 = 0 $ 
and the normal to this surface  has direction
\be
N_{\mu} =\nabla_{\mu}f= \delta_{\mu 1} \,;
\ee
its norm
\be
N_c N^c= g^{\mu\nu}\delta_{\mu 1}\delta_{\nu 1}=   g^{rr}
= \frac{1}{2} \left( 1-\frac{2\Lambda r^2}{3}\right)  \label{Norm}
\ee
is positive for any $r<r_*=\sqrt{\frac{3}{2\Lambda} }$. Taking the limit 
$r\rightarrow 0^{+}$ in this equation, one obtains $N^c  N_c \Big|_{r=0}= 
1/2$, hence $N^c$ is spacelike and the central singularity at $r=0$ is 
timelike.

In the $\Lambda\rightarrow 0$ limit to Wyman's ``other'' solution, 3-spaces 
of constant time are infinite, the coordinate $r$ extends to infinity, and 
the geometry describes a naked singularity embedded in a spacetime which is 
not asymptotically flat because the corresponding 
Ricci tensor
\be
{\cal R}_{ab} = \kappa \nabla_a \phi \nabla_b \phi  
= \kappa \phi_0^2 \delta_{a0} \delta_{b0} 
\ee
does not vanish as $r\rightarrow +\infty$ and the energy density $\rho 
\sim 1/r^2$ diverges when integrated between a finite radius and 
infinity (see below).

\subsection{Quasilocal mass}

In spherical symmetry, the Misner-Sharp-Hernandez 
mass  $M_\text{MSH}$ contained in a ball of radius $r$ is 
defined by \cite{MSH1,MSH2} 
\be
 1-\frac{2GM_\text{MSH}}{r} \equiv \nabla^{c} r \nabla_{c}r \,,
\ee
where $r$ is the areal radius. The Hawking-Hayward quasilocal mass 
\cite{Hawking:1968qt, Hayward:1993ph}  reduces to the 
Misner-Sharp-Hernandez mass in spherical symmetry and is the Noether charge 
associated with the conservation of the Kodama current 
\cite{Hayward:1994bu}. For the Sultana-Wyman solution with 
$\Lambda>0$, the Misner-Sharp-Hernandez mass is \cite{Banijamali:2019gry}
\begin{equation}
M_\text{MSH}(r)=\frac{r}{4G}\Big(1+\frac{2\Lambda r^{2}}{3}\Big) 
\,.\label{eq:ecce}
\end{equation}
By comparison, the Misner-Sharp-Hernandez mass of a ball in de Sitter 
space is $M_\text{dS}(r) = \frac{ \Lambda r^3}{6G} $, so the scalar field 
$\phi$ 
in the Sultana-Wyman geometry contributes an amount $ r/(4G) $ added to 
the mass of de Sitter space. More precisely, the mass in 
Eq.~(\ref{eq:ecce}) splits as
\be
M_\mathrm{MSH} (r)= \frac{4\pi r^3}{3} \, \frac{\Lambda}{\kappa} + \frac{4\pi 
r^3}{3} \, \frac{1}{2\kappa \, r^2} = 
\frac{4\pi r^3}{3} \left( \rho_{(\phi)} + \rho_{\Lambda} 
\right)  \,,
\ee
where we used Eq.~(\ref{rho1}).

For $\Lambda=0$ (in which case $0\leq r <+\infty$), the mass 
$M_\text{MSH}(r)$ of the Wyman solution diverges linearly  
as the areal radius $r\rightarrow +\infty$, showing again that this 
geometry is not asymptotically flat (in which case $M_\mathrm{MSH}$ would be 
finite \cite{Faraoni:2020mdf}).

For $\Lambda>0$, the total Misner-Sharp-Hernandez mass  contained in  
the finite Sultana-Wyman slices of constant time is
\begin{equation}
M_\text{MSH}\left( r=\sqrt{\frac{3}{2\Lambda}} \right)=\sqrt{ 
\frac{3}{8\Lambda G^2}} \,. 
\end{equation}

\section{The Sultana-Wyman geometry as a finite fluid ball: Buchdahl-Land 
and Iba\~nez-Sanz}
\label{sec:3}
\setcounter{equation}{0}

It is well known that a minimally coupled scalar field is equivalent to a 
perfect fluid and that a free scalar corresponds to a stiff fluid with 
equation of state $P=\rho$. Therefore, the Sultana-Wyman solution can be 
interpreted as describing a spacetime filled with a stiff fluid and a 
cosmological constant. Indeed, it corresponds to a special case of a 
previous stiff fluid solution. This fact was apparently unknown to Wyman 
in the case $\Lambda=0$, but Sultana identifies his generalization of 
Wyman's solution to $\Lambda>0$ with a solution generated by Iba\~nez \& 
Sanz \cite{IbanezSanz} using the Heintzmann technique \cite{Heintzmann69}. 
Iba\~nez \& Sanz correctly identify it with the previous Buchdahl-Land 
solution \cite{BuchdahlLand} which is, in turn, a special case of the 
Tolman~IV class of solutions of the Einstein equations with $\Lambda$ 
\cite{Tolman39}. The Buchdahl-Land solution was recently used by Jowsey \& 
Visser \cite{Jowsey:2021ixg} as an example in an unrelated context, the 
question of the existence of a maximum force in general relativity.

The Tolman~IV solution of the Einstein equations with $\Lambda$ and a 
perfect fluid  is \cite{Tolman39} 
\begin{eqnarray}
ds^2 &=& -\left(1+ \frac{r^2}{A^2} \right) dt^2 
+\frac{1+2r^2/A^2}{\left( 1-\frac{r^2}{R^2}\right) 
\left(1+\frac{r^2}{A^2} \right) }  \,  dr^2  \nonumber\\
&&\nonumber\\
&\, & +r^2 d\Omega^2_{(2)} \,,
\end{eqnarray}
where $A$ and $R$ are constants. By using the dimensionless time $ 
\tau\equiv t/A$, this line element is rewritten as
\begin{eqnarray}
ds^2 &=& -\left( A^2 +r^2\right) d\tau^2 +\frac{ A^2 
+2r^2}{\left(1-r^2/R^2\right)\left(A^2 +r^2 \right)} \, dr^2 
\nonumber\\
&&\nonumber\\
&\, & +r^2 d\Omega_{(2)}^2 \,.
\end{eqnarray}
Taking the limit in which the parameter $A\rightarrow 0$ yields
\be
ds^2 = -r^2 d\tau^2 +\frac{ 2}{1-r^2/R^2} \, dr^2 +r^2 
d\Omega_{(2)}^2 \,.
\ee
Redefining the time coordinate as $ \tau \equiv \sqrt{\kappa}\, 
\bar{t}$ and identifying $2\Lambda/3 \equiv 1/r_\text{H}^2$ with $1/R^2$, 
one obtains the Buchdahl-Land line element \cite{BuchdahlLand}
\be
ds^2 = -\kappa\, r^2 d\bar{t}^2 +\frac{ 2}{1-2\Lambda r^2/3} \, dr^2 +r^2 
d\Omega_{(2)}^2 \,,
\ee
which coincides with the Sultana-Wyman solution~(\ref{geo}). In this 
notation, the latter has its boundary at radius $ \sqrt{3/(2\Lambda)} 
= R $.

The scalar field is redefined according to 
\be
\phi=\phi_0 \, t = \phi_0 A\tau = \phi_0 \sqrt{\kappa}\, \bar{t} \equiv 
\bar{\phi}_0 \, \bar{t} \,.
\ee

Let us relate the ``standard'' view in the literature (the 
Buchdahl-Land/Iba\~nez-Sanz \cite{BuchdahlLand,IbanezSanz} geometry as a 
stiff fluid GR solution) and the Sultana-Wyman view of the same geometry 
as a scalar field solution with $\Lambda>0$. Starting from the latter, the 
expression of the scalar field stress-energy tensor \cite{Waldbook}
\be
T_{ab}^{(\phi)} = \nabla_a \phi \nabla_b \phi -\frac{1}{2} \, g_{ab} 
\nabla^c \phi\nabla_c \phi -V g_{ab}
\ee
gives the well-known energy density and pressure
\begin{eqnarray}
\rho_{(\phi)} &=& -\frac{1}{2} \,\nabla^c \phi\nabla_c \phi +V(\phi) \,,\\
&&\nonumber\\
P_{(\phi)} &=& -\frac{1}{2} \,\nabla^c \phi\nabla_c \phi -V (\phi) \,,
\end{eqnarray}
which make it clear that a free scalar field 
corresponds to a stiff fluid with equation of state 
$P_{(\phi)}=\rho_{(\phi)}$. 

If one regards the Sultana-Wyman solution as a free scalar field solution 
of the Einstein equations with cosmological constant $\Lambda >0$, then 
using $\phi= \bar{\phi}_0 \, \bar{t} $, one has 
\be
\rho_{(\phi)} = P_{(\phi)} = \frac{ \bar{\phi}_0^2}{2\kappa \, r^2} \,,
\ee
but the {\em total} effective energy density and pressure are 
obtained by viewing 
the $\Lambda$-term as an effective fluid with stress-energy tensor 
$T_{ab}^{(\Lambda)}=-\frac{\Lambda}{\kappa} \, g_{ab}$ in the right hand 
side of the Einstein equations,
\begin{eqnarray}
\rho_\text{tot} &=& \frac{ \bar{\phi}_0^2}{2\kappa \, r^2} 
+\frac{\Lambda}{\kappa} = \frac{1}{16\pi} \left( \frac{ 
\bar{\phi}_0^2}{r^2} + \frac{3}{R^2} \right) \,, \label{questa}\\
&&\nonumber\\
P_\text{tot} &=& \frac{ \bar{\phi}_0^2}{2\kappa \, r^2} 
-\frac{\Lambda}{\kappa} = \frac{1}{16\pi} \left( \frac{ 
\bar{\phi}_0^2}{r^2} 
-\frac{3}{R^2} \right) \,,\label{quella}
\end{eqnarray}
where we used the fact that $\Lambda=\frac{3}{2R^2}$. Alternatively, one 
can regard the Sultana-Wyman spacetime as a solution of the Einstein 
equations without cosmological constant but with a scalar field in the 
constant potential $V(\phi)=\Lambda/\kappa$, with the same result. If we 
set 
$\bar{\phi}_0=1$, Eqs.~(\ref{questa}) and (\ref{quella}) match Eqs.~(3.60) 
of Jowsey \& Visser \cite{Jowsey:2021ixg}, who do not contemplate scalar 
fields and view the Buchdahl-Land solution as a stiff fluid ball.

The energy density and pressure are singular as $r\rightarrow 0^{+}$ and 
the pressure $P_{(\phi)}$ vanishes at the radius 
\be
\frac{ | \bar{\phi}_0| R}{\sqrt{3}} = \frac{ |\bar{\phi}_0| r_\text{H} 
}{\sqrt{3}} 
\ee
the radius that is usually taken as the boundary of the star in the 
literature (this corresponds to $R_s \equiv R/\sqrt{3}$ in 
\cite{Jowsey:2021ixg}).

Keeping $\bar{\phi}_0$ general, we encounter two possible situations:

\begin{itemize}

\item If $ |\bar{\phi}_0 | <\sqrt{3} $, the boundary $r_*$ of the star  is 
below the Sultana-Wyman maximum radius, $ r_* < r_\text{H} $;

\item If $ |\bar{\phi}_0 | =\sqrt{3} $ the fluid configuration fills the 
entire Sultana-Wyman 3-space, {\em i.e.}, it is not a star.

\end{itemize}

These scenarios are discussed in the following.

\subsection{Star boundary below $\sqrt{ \frac{3}{2\Lambda}} $}

Assuming that the star extends from the origin $r=0$ (where, however, 
there is spacetime singularity---see the discussion below) to a boundary 
$r_0$, one has to match the interior Sultana-Wyman scalar field solution 
with an exterior in order to build a stellar model. In spherical stellar 
models, the standard practice consists of matching an interior fluid 
solution with a Schwarzschild exterior ({\em e.g.}, \cite{Delgaty:1998uy, 
Faraoni:2021nhi}). However, having established that the BLSWIS geometry 
solves the Einstein equations with $\Lambda>0$, the interior must be 
matched with a Schwarzschild-de Sitter/Kottler exterior, which is the 
unique solution in this case, according to a straightforward 
generalization of the Birkhoff theorem 
\cite{Schleich:2009uj,Faraoni:2017uzy}.

There are two possibilities: either one regards the interior as a stiff 
fluid solution of the Einstein equations (with $\Lambda>0$), or as a free 
scalar field solution of the Einstein equations (with $\Lambda >0$). In 
the first case, since the pressure goes to zero at the star boundary $r_0< 
\sqrt{ \frac{3}{2\Lambda}} $, one would be tempted to match with a 
Schwarzschild exterior, as done for all fluid models of 
stars\footnote{ Buchdahl \& Land \cite{BuchdahlLand}, Iba\~nez \& Sanz 
\cite{IbanezSanz}, and Jowsey \& Visser \cite{Jowsey:2021ixg} do not 
discuss this matching nor refer to it, but it is implicit in the large 
literature on stellar models that a stellar interior must be matched with 
an exterior Schwarschild \cite{Stephanietal}.} ({\em e.g.}, 
\cite{Delgaty:1998uy}), but Schwarzschild is not a solution of the 
Einstein equations in a $\Lambda>0$ vacuum. Therefore, the interior should 
be matched smoothly with a Schwarzschild-de Sitter exterior, but this is 
impossible because the interior pressure $P_\mathrm{tot}$ given by 
Eq.~(\ref{quella}) is always larger than the exterior pressure 
$P_\Lambda=-\Lambda/\kappa <0 $. One could allow for a discontinuity of 
matter on the star boundary, but this implies the presence of a layer of 
material on that surface, which is not a physical model of a star.

Let us consider the second possibility. Since the scalar field $\phi(t)$ 
does not depend on the spatial coordinates, it cannot be set to zero at 
the star boundary, or to a constant (with respect to time) in the star 
exterior, therefore the exterior solution must also be a scalar field 
solution of the Einstein equations with $\Lambda\neq 0$. The fluid ball is 
not surrounded by vacuum and its exterior geometry cannot be 
Schwarzschild-de Sitter/Kottler \cite{Schleich:2009uj,Faraoni:2017uzy}. 
The homogeneous scalar field is linear in time, a feature that persists in 
the exterior by continuity.  Therefore, the interior Sultana-Wyman 
solution does not match to the Fisher geometry \cite{Fisher:1948yn} 
either, for which $\phi = \phi(r)$.

This situation is rather curious: in the two cases above, the field 
equations are different, and it happens that a certain geometry solves 
both.\footnote{This situation is quite common: for example, any physically 
reasonable theory of gravity admits the 
Friedmann-Lema\^itre-Robertson-Walker solution. In these situations, 
although the field equations are very different 
\cite{Clifton:2011jh,Faraoni:2010pgm}, the same geometry solves both.} In 
the present problem with the stiff fluid solution {\em of the $\Lambda>0$ 
Einstein equations}, one wants to cut the solution at the specific value 
$r_*$ of radius where $P$ vanishes and join it smoothly with an exterior 
solution. However, in the other interpretation in which the geometry is a 
scalar field solution of the Einstein equations {\em with cosmological 
constant $\Lambda > 0 $}, the field equations are different and one should 
not expect {\em a priori} that joining smoothly the same interior geometry 
with an exterior one is possible, or physically meaningful, or that it 
gives the same result. The interior solution solves two different sets of 
field equations, but continuing it smoothly to an exterior is an issue. 
The two different points of view contemplating different sources with a 
(hidden) $\Lambda$ behave differently with respect to the continuation to 
an exterior. In one case, a discontinuous matching with a Schwarschild-de 
Sitter exterior is the only possibility, which entails a layer of material 
at the star surface. In the other situation, one must match the same 
interior geometry with an exterior that has homogeneous scalar field and 
$\Lambda>0$ and is not Schwarzschild-de Sitter, or else one must impose 
the additional unphysical requirement that the scalar field is 
discontinuous. None of these two situations is interesting to build a 
physical model of a relativistic star (with or without $\Lambda$).

In any case, because of the central singularity, there is another 
potentially very serious issue in interpreting the BLSWIS solution as 
describing a fluid ball. In the literature, various authors seem to 
content themselves with assuming that only regions with $r>0$ of this 
Buchdahl-Land solution describe realistic star geometries (see the 
comments by Buchdahl \& Land \cite{BuchdahlLand}, Iba\~nez \& Sanz 
\cite{IbanezSanz} and Jowsey \& Visser \cite{Jowsey:2021ixg} to this 
regard). In their monumental review of exact GR solutions, Stephani, 
Kramer, MacCallum, Hoenselaers \& Herlt also suggest using solutions with 
a central singularity to model the outer layers of composite spheres 
\cite{Stephanietal}, and using regions with different equations of state 
is common in the modelling of Newtonian stars, when their interiors are 
not well mixed. However, selecting certain limited spacetime regions as 
realistic solutions ultimately involves further matching with other 
non-singular solutions extending down to $r=0$. To the best of our 
knowledge, this possibility is not actively explored in the literature.

\subsection{Star boundary at  $r=\sqrt{ \frac{3}{2\Lambda}} $}

This potential possibility corresponds to a very strange situation. The 
pressure becomes negative in the region $ \frac{ | \bar{\phi}_0| 
R}{\sqrt{3}} < r< r_*$ (while $\rho_{(\phi)}$ remains positive), which is 
unphysical for a stellar interior. An exterior must necessarily have the 
same value of the cosmological constant $\Lambda>0$ and the same scalar 
field $\phi=\phi_0 t$---that is, the solution is again Sultana-Wyman, but 
we know that its 3-spaces of constant time have finite extension, 
therefore one cannot consider an ``exterior''.  Attempts to describe a 
stellar configuration with boundary at $r=\sqrt{ \frac{3}{2\Lambda}} $ 
seem doomed.

\section{Conclusions}
\label{sec:4}
\setcounter{equation}{0}

The history of the BLSWIS solution of the Einstein equations is a bit 
convoluted: it is derived either as a solution with a free homogeneous scalar 
field $\phi(t)$ and cosmological constant $\Lambda>0$ \cite{Sultana:2015lja}, 
or as special limits of interior solutions for a relativistic star with a 
perfect fluid and, superficially, $\Lambda=0$ \cite{IbanezSanz, BuchdahlLand, 
Tolman39}. However, when it is obtained through these special limits, there 
is a positive cosmological constant hidden in this solution that was not 
evident in the more general fluid solutions. The crucial difference between a 
real fluid and a cosmological constant, even when the latter is treated as an 
effective fluid, is that the former can be confined to a limited region of 
spacetime and vanish outside of it, but the latter permeates all of 
spacetime.

While the spacetime geometry is the same in the Sultana-Wyman and the 
perfect (stiff) fluid solutions, the boundary conditions at the surface of 
the would-be star differ in the two contexts. Matching smoothly the BLSWIS 
``interior'' to an ``exterior'' in order to build a stellar model 
(necessarily, for $\Lambda>0$ and with a homogeneous scalar field) does 
not make much sense physically, as discussed above. In particular, the 
constant time slices of the BLSWIS geometry are finite. In the 
Oppenheimer-Snyder model of gravitational collapse \cite{Datt38, 
Oppenheimer:1939ue}, a finite Friedmann-Lema\^itre-Robertson-Walker 
universe with positively curved spatial sections and filled with dust is 
matched 
smoothly with a Schwarschild exterior, and this case bears some 
resemblance to the BLSWIS case. However, in order to match to the 
Schwarzschild exterior, the fluid in the interior must necessarily 
have zero pressure. In the BLSWIS case, an exterior must necessarily have 
the same value of the cosmological constant $\Lambda>0$ and the same 
scalar field $\phi=\phi_0 t$---that is, the solution is again 
Sultana-Wyman, but we know that its 3-spaces of constant time have finite 
extension, therefore one cannot consider an ``exterior''.

When examining the other boundary at the star's centre $r=0$, the central 
singularity of the BLSWIS geometry does not bode well for using the Wyman 
geometry or its Sultana generalization to describe the interior of stars. 
Iba\~nez \& Sanz \cite{IbanezSanz} and also Jowsey \& Visser 
\cite{Jowsey:2021ixg} join the existing literature \cite{Stephanietal} in 
regarding this geometry as capable of describing certain regions of 
relativistic stars. While this may be the case, extra caution must be 
exerted when applying this GR solution to realistic situations. Overall, 
the BLSWIS solution of the Einstein equations does not lend itself  
to model physically meaningful relativistic stars and joins the graveyard 
of exact solutions of the Einstein equations originally meant to model 
fluid balls which fail to do so for one reason or another 
\cite{Delgaty:1998uy}. The lesson learned from the BLSWIS case is that, in 
order to build realistic models of relativistic stars (or of regions of 
them), it is not sufficient to solve analytically the Einstein equations 
with a perfect fluid source, but attention must be paid to the exterior 
and to the star boundary, since a cosmological constant or a inhomogeneous 
matter source cannot be eliminated when passing from the  interior 
to the exterior.

\begin{acknowledgments} 

We are grateful to Andrea Giusti for a discussion. This work is supported, 
in part, by the Natural Sciences \& Engineering Research Council of Canada 
(Grant~2016-03803 to V.F.) and by a Bishop’s University Graduate Entrance 
Scholarship to S.J.

\end{acknowledgments}

\appendix
\section{Klein-Gordon equation for the Sultana-Wyman spacetime}
\renewcommand{\theequation}{A.\arabic{equation}}
\label{Appendix:A}

It is straightforward to show that $\phi=\phi_0 t $ solves the 
Klein-Gordon equation~(\ref{KleinGordon}), which reads
\be
\Box \phi = \frac{1}{ \sqrt{-g}} \, \partial_{\mu} \left( \sqrt{-g} \, 
g^{\mu\nu} \partial_{\nu} \phi \right) =0 
\ee
for both the Wyman solution (for which $\Lambda=0$) and for its Sultana 
generalization with $\Lambda>0$.

Using $\sqrt{-g} = \sqrt{ \frac{2\kappa}{1-2\Lambda r^2/3}} \, r^3 
\sin\vartheta $ and $\partial_{\mu}\phi = \phi_0 \delta_{\mu 0}$, this 
equation reduces to
\be
\partial_t \left( \frac{ \phi_0 }{ \sqrt{\kappa} } \, \frac{ 
\sqrt{2}\, r \sin\vartheta}{\sqrt{ 1-2\kappa \, r^2/3}} \right)=0 \,,
\ee
which is trivially satisfied since the argument of the round bracket 
depends only on $r$.

When the cosmological constant $\Lambda>0$ is included by Sultana in the 
picture, it is equivalent to regard the total matter content of 
spacetime as 1)~a free 
scalar field $\phi$ with $\Lambda$ in the Einstein equations, or 2)~as a 
scalar field with the constant potential $V=\Lambda/\kappa$ and no 
cosmological constant in the Einstein equations. In the first case, the 
Klein-Gordon equation does not change in form. By making the second 
choice, the  Klein-Gordon equation 
for $\phi$ would be modified by the potential $V(\phi)$ according to 
\be
\Box \phi -\frac{dV}{d\phi}=0 \,,
\ee
but since $dV/d\phi \equiv 0$ for $V=\Lambda/\kappa$, the form of this 
equation 
is unchanged (however, the cosmological constant $\Lambda$ or, 
alternatively, the potential $V$ changes the field 
equations~(\ref{Einstein}) for the metric and, accordingly, the solution 
changes from the Wyman to the Sultana-Wyman one.

\end{document}